\documentclass[preprint,showpacs,preprintnumbers,amsmath,amssymb]{revtex4}

\usepackage{graphicx}
\usepackage{dcolumn}
\usepackage{bm}
\usepackage{amsfonts}%
\usepackage{amsmath}

\begin{document}


\title{Dissociative recombination of  H$_3^+$ in the ground and excited vibrational states.}
\author{Samantha Fonseca dos Santos$^\dag$, Viatcheslav Kokoouline$^\dag$, and Chris H. Greene$^\ddag$}
\affiliation{$^\dag$ Department of Physics University of Central Florida, Orlando, FL 32816, USA\\ $^\ddag$ Department of Physics and JILA, University of Colorado, Boulder, Colorado 80309-0440, USA}
\date{\today}

\begin{abstract}
The article presents calculated dissociative recombination (DR) rate coefficients for H$_3^+$. The previous theoretical work on H$_3^+$ was performed using the adiabatic hyperspherical approximation to calculate the target ion vibrational states and it considered just a limited number of ionic rotational states. In this study, we use accurate vibrational wave functions and a larger number of possible rotational states of the H$_3^+$ ground vibrational level. The DR rate coefficient obtained is found to agree better with the experimental data from storage-ring experiments than the previous theoretical calculation. We present evidence that excited rotational states could be playing an important role in those experiments for collision energies above 10 meV. The DR rate coefficients calculated separately for ortho- and para-H$_3^+$ are predicted to differ significantly at low energy, a result consistent with a recent experiment. We also present DR rate coefficients for vibrationally-excited initial states of H$_3^+$, which are found to be somewhat larger than the rate coefficient for the ground vibrational level.
\end{abstract}

\pacs{34.80.Ht 34.80.Kw 34.80.Lx }
\vspace*{0cm}

\maketitle

\section{Introduction}

Dissociative recombination (DR) of the simplest polyatomic ion H$_3^+$ 
\begin{equation}
\mathrm{H}_3^+ + e^- \longrightarrow \mathrm{H}_2+\mathrm{H}\ \mathrm{or}\ \mathrm{H}+\mathrm{H}+\mathrm{H}
\end{equation}
has been studied for several decades both in experiment and theory \cite{oka_general,larsson00,kokoouline01,kokoouline03b}. The measured rate of the reaction is relatively fast \cite{larsson00,sundstrom94,jensen01,larsson93,mccall03,mccall04,kreckel02,kreckel05} for electron energies below 1 eV, which was eventually attributed to the strong Jahn-Teller coupling between vibrational motion of the ion and the incident $p$-wave in the electronic continuum \cite{kokoouline01,kokoouline03b,kokoouline03a}. Currently, there is general agreement between theory \cite{kokoouline03b,kokoouline03a,orel93} and most recent experiments \cite{sundstrom94,larsson00,jensen01,larsson93,mccall03,mccall04,kreckel02,kreckel05} for the DR rate coefficient in H$_3^+$. However, the detailed energy dependence of the theoretical rate coefficient \cite{kokoouline03b,kokoouline03a} in the range 0-2 eV exhibits differences from the rate coefficient measured in recent high resolution storage ring experiments \cite{mccall03,mccall04,kreckel05}. One prominent point of disagreement is the much more pronounced resonance structure in the theoretical rate coefficient; this plethora of resonances is associated with Rydberg states of the neutral molecule H$_3^*$. Although similar resonance structure is visible in the experimental data, it is much less pronounced.

Dissociative recombination of H$_3^+$ is a four-body problem: Because the DR process starts with an electron-H$_3^+$ collision, one must account for the motion of the electron and its coupling with the molecular degrees of freedom. In order to represent Jahn-Teller coupling between electronic and vibrational motion, one must take into account at least two degrees of freedom of vibrational motion (two hyperangles in our approach). Finally, the third vibrational coordinate (the hyper-radius) should be included in order to describe the dissociative channel. Therefore, the theoretical approach includes several ingredients \cite{kokoouline03b,kokoouline03a} and a number of approximations were made to simplify the calculations in our previous theoretical study \cite{kokoouline03b,kokoouline03a}.

One of the important approximations used in Refs.  \cite{kokoouline03b,kokoouline03a} is the adiabatic hyperspherical approximation: motion in the hyper-radial coordinate was treated as adiabatic compared to motion in the hyperangles, i.e. motion in the hyperangles was considered to be much faster than motion in the hyper-radius. Correspondingly, vibrational states $\Phi_v$ of the ion have been represented as simple products of hyperangular and hyper-radial functions. Somewhat surprisingly, this approximation describes reasonably well (at about the 1\% level) vibrational energies of the ion for the low vibrational states in low-lying hyperspherical potential curves (see Table \ref{table:vibr_energies}). For excited vibrational states with energies higher than 1 eV above the ground rovibrational level, the error increased to the vicinity of 10 meV. Such highly excited vibrational states may support Rydberg states of the neutral H$_3^*$ molecule in the energy region corresponding to the low-energy DR process under consideration. Accordingly, the absolute positions of such Rydberg states were presumably calculated with an error of about 10 meV. We should also mention that the highest vibrational levels that must be included in our theoretical treatment (in order to represent the initial dissociation of the neutral molecule) have around 4 eV of vibrational energy. Another source of error in the position of Rydberg states is the neglected energy-dependence of quantum defects used in Refs.  \cite{kokoouline03b,kokoouline03a}. Analysis of an {\it ab initio} calculation \cite{mistrik00} shows that the quantum defects are in fact only weakly energy-dependent close to the equilibrium geometry of the molecular ion. The maximum error in positions of Rydberg states for $n\approx 2-3$ associated with this approximation is estimated to be around 10-15 meV, at least for Rydberg states close to that equilibrium configuration. Certain Rydberg states appear in the DR spectrum as resonances and, therefore, play an important role in the detailed comparison of the theoretical DR rate coefficient with the high-resolution experimental data. On the other hand, the DR rate coefficient that has been thermally averaged over a Maxwell-Boltzmann distribution should be comparatively insensitive to the detailed positions of Rydberg states, as long as the average level density and the resonance widths are described correctly. Indeed, the theoretical thermal rate coefficient calculated in the previous study \cite{kokoouline03b} agrees with the experimentally-measured thermal rate coefficient.

The adiabatic hyperspherical approximation implemented previously for the ionic vibrational eigenstate calculation neglected all non-adiabatic coupling between different adiabatic channels.  This approximation could also adversely affect the calculated DR rate coefficient since the main DR mechanism in H$_3^+$ is indirect: neglect of some non-adiabatic effects might therefore be expected to cause an underestimation of the theoretical DR rate coefficient. However, the most important coupling responsible for the high DR rate coefficient of H$_3^+$ is non-Born-Oppenheimer Jahn-Teller coupling, which has been accounted for in the previous study. 

Finally, in the previous study only a few rotational levels of the initial state of the ion were included up to $N^+=3$. (The rotational state (3,1) was not included. Here and below we use the generally accepted notations for rotational states of the ion \cite{lindsay01}, where the two numbers in () are the total angular momentum $N^+$ of the ion and its projection $K^+$ on the ionic symmetry axis.) The inclusion of a larger number of rotational states has since been recognized as possibly important due to the following: If the electron energy is high enough, there is a large probability that the H$_3^+$ ion is excited into a higher rotational level \cite{kokoouline03b,faure06}. In fact, the probability of this rotational excitation is high enough to be competitive with the DR process. Thus, for electron energies above 10 meV, higher rotational states of the ion could be populated in the storage ring even though, initially, the ionic beam has been prepared in the ground rovibrational state. Excitations can happen, for instance, when the ions pass through the toroidal region or the electron cooler \cite{mccall04}.

The present study improves upon two of the three approximations discussed above. First, we calculate accurate vibrational states of the ion instead of relying on the adiabatic hyperspherical approximation; the adiabatic hyperspherical {\it representation} is still utilized, but since channel coupling is now included, the calculated eigenspectrum can be made arbitrarily accurate, in principle. Second, we account for a larger number of initial rotational states. Finally, we calculate the rate coefficients of DR processes that start from two different excited vibrational states of H$_3^+$.

The article is organized as follows. Section II briefly summarizes our theoretical approach and, in particular, the method to obtain accurate vibrational energies and wave functions of the ion. Section \ref{sec:averaging} discusses the various averaging procedures that must be performed on the raw theoretical DR rate coefficient in order to compare theory with the data from existing storage ring experiments. Section \ref{sec:results} presents our results and Section \ref{sec:conclusion} gives our conclusions.

\section{Theoretical approach}

The procedure of calculation of the cross-section is very similar to that described in Ref. \cite{kokoouline05}.  Here we briefly summarize the main steps. 

The treatment is based on construction of the multi-channel electron-H$_3^+$ scattering matrix $\cal S$. The asymptotic channels $|i\rangle$ of the matrix corresponds to different rovibrational levels of the ion. After a collision with the electron the rovibrational state of the ion may change:
\begin{equation}
\label{eq: collision process}
e^- + {\rm H}_{3}^{+}(i) \longrightarrow e^- + {\rm H}_{3}^{+}(i')\,.
\end{equation}
However, the collision conserves the overall symmetry of the system.  (That is, $\cal S$ is diagonal in the irreducible representation $\Gamma_{tot}$ of the symmetry group $D_{3h}$ of the Hamiltonian.) We assume that the electronic state of the ion is the ground $^1A_1'$ state. Since protons are fermions, for H$_3^+$ in the ground electronic state, the allowed irreducible representations of the total nuclear wave function (including space and nuclear spin coordinates) are $A_2'$ and $A_2''$ of the $D_{3h}$ symmetry group.  When the scattering matrix ${\cal S}_{i,i'}$ is constructed, highly excited vibrational levels of H$_3^+$ are included. These vibrational levels in fact represent the discretized vibrational continuum, and they have finite lifetimes with respect to dissociation. The energy of the incident electron is not high enough for these levels to represent open channels for dissociative ionization, but the electron can be captured into the Rydberg states attached to the highly excited levels. If this happens, the system dissociates instead of ionizing, because such low principle quantum number Rydberg states are locally open for dissociation. This causes the electron-ion scattering matrix $S$ to be non-unitary, and the `defect' from unitarity of the relevant columns of $S$ can be identified with the dissociation probability.

The total wave function of the ion-electron system is constructed by taking into account all symmetry restrictions determined by the two allowed irreducible representations of the system, which is discussed in detail in Refs.\cite{kokoouline03b,kokoouline03a}. The wave functions of the $A_2'$ and $A_2''$ irreducible representations of the channel states $|i\rangle$ are constructed from the product of the rotational, vibrational, nuclear, and electronic degrees of freedom of the system (see also Eqs. (2) of Ref. \cite{kokoouline03a}):
\begin{eqnarray}
\label{eq:Total Wave Function}
|i\rangle=\hat P^{sym}\Phi_{total}\,,\nonumber\\
\Phi_{total} = \Phi_{rot}\Phi_{vib}\Phi_{ns}\Phi_{el}\,.
\end{eqnarray}
The operator $\hat P^{sym}$ projects the product $\Phi_{total}$ on the corresponding irreducible representation, $A_2'$ or $A_2''$.  Each factor in the second equation is calculated by the diagonalization of the respective Hamiltonian, except $\Phi_{el}$. In the above equation, $\Phi_{rot}$ is the rotational wave function of the ion, $\Phi_{vib}$ is the vibrational wave function, $\Phi_{ns}$ is the nuclear spin wave function, and $\Phi_{el}$ represents the wave function of the incoming electron. The functions $\Phi_{el}$ do not diagonalize completely the clamped-nucleus electronic Hamiltonian; the matrix of the electronic Hamiltonian represented by  $\Phi_{el}$ has non-diagonal elements responsible for Jahn-Teller coupling in H$_3$ \cite{mistrik00,staib90a,staib90b,stephens94,stephens95}.

The rotational, nuclear spin, and electronic functions in the product $\Phi_{total}$ are represented in the same way as in our previous study, Ref. \cite{kokoouline03b}. But the vibrational part $\Phi_{vib}$ is obtained differently here, because we no longer utilize the hyperspherical adiabatic approximation \cite{Macek68, Fano_review, Lin_review}. In that approximation, the non-adiabatic couplings between different hyperspherical adiabatic states of the same vibrational symmetry were entirely neglected, an approximation that works reasonably well for low vibrational levels of H$_3^+$ ion. In this study we use much better vibrational states calculated using the slow variable discretization (SVD) approach \cite{tolstikhin96,kokoouline2006}, where non-adiabatic couplings are taken into account. Table \ref{table:vibr_energies} compares accuracy of calculations of the present approach with the hyperspherical adiabatic approach. The calculations of H$_3^+$ vibrational states by Jaquet {\it et al.} \cite{jaquet98}, obtained for the same Born-Oppenheimer ionic surface that we use here \cite{cencek98}, are taken as an `exact' reference.

Once $\Phi_{total}$ are known, the matrix elements ${\cal S}_{i,i'}$ are calculated from the scattering matrix $S_{\Lambda,\Lambda'}(\cal Q)$ depending on three distances between protons in H$_3^+$ and describing the $e^-$+H$_3^+$ collision in the molecular frame, where the appropriate quantum numbers are projection $\Lambda$ of the electronic orbital momentum on the ionic principal axis and the set of internuclear coordinates $(\cal Q)$. As in the previous study \cite{kokoouline03b,kokoouline03a}, we consider only the `$p-$wave' of the electron, when it moves beyond the range of the ionic core. The nonspherical nature of the electron-ion interaction potential undoubtedly mixes other electronic orbital momenta when the electron is at short range, but scattering calculations have shown that the probability for an incident $p-wave$ electron to scatter into an $s$ or $d$ orbital momenta is quite low in this near-threshold energy range. The matrix $S_{\Lambda,\Lambda'}(\cal Q)$ is obtained from the reaction matrix $K^0_{\Lambda,\Lambda'}(\cal Q)$ given by formulas in Refs. \cite{staib90a,staib90b}. As was mentioned above, the electronic Hamiltonian and, correspondingly, the matrices related to body-frame scattering, i.e. $S$ and $K$, are not diagonal. The nonzero off-diagonal elements $K^0_{1,-1}$/$S_{1,-1}$ are due to the Jahn-Teller coupling.

When the total energy of the $e^-$+H$_3^+$ system is not high enough for all the channels $|i\rangle$ to be energetically open for ionization, the usual situation, the physically meaningful scattering matrix ${\cal S}^{phys}(E)$ is obtained from ${\cal S}_{i,i'}$ by the  standard closed-channel elimination procedure of multi-channel quantum defect theory (MQDT) \cite{seaton_review,aymar96}. The dissociative recombination rate coefficient is then calculated using the unitarity `defect' of the corresponding columns of ${\cal S}^{phys}$ \cite{kokoouline03b,kokoouline03a}. In order to compare our results with the data from storage ring experiments, we carry out a number of averaging procedures in order to model the experimental conditions, as detailed below.

\begin{table}
\vspace{0.3cm}
\begin{tabular}{|p{2.cm}|p{4cm}|p{4cm}|p{4cm}|}
\hline
$v_1v_2^{l_2}$, irrep.&adiab. approx.     &SVD calc.     &Jaquet {\it et al} \cite{jaquet98}\\
\hline
$00^0\,A_1$    &  0            &  0            &    0\\
$10^0\,A_1$    &  3188            &  3177.5        &    3178.15    \\
$02^0\,A_1$    &  4754            &  4777.9        &    4778.01    \\
$20^0\,A_1$    &  6273            &  6260.6        &    6261.81    \\
$03^0\,A_1$    &  7382            &  7275.8          &     7285.32    \\
$12^0\,A_1$    &  7648            &  7772.9        &     7769.06    \\
$04^0\,A_1$    &  8979            &  8995.6        &     9000.58    \\
$30^0\,A_1$    &  9248            &  9255.5        &     9252.08    \\
$13^3\,A_1$    &  10129        &  9958.6        &     9963.98    \\
$22^0\,A_1$    &  10420        &  10598.        &     10590.51\\
$05^3\,A_1$    &              &  10912.        &     10915.47\\
\hline
$01^1\,E$    &  2516            &  2521.1        &    2521.20\\
$02^2\,E$    &  5001            &  4996.6        &    4997.73\\
$11^1\,E$    &  5554            &  5552.9        &    5553.95\\
$03^1\,E$    &  6978            &  6999.2        &    7005.81\\
$12^2\,E$    &  7897            &  7865.2        &    7869.82\\
$21^1\,E$    &  8478            &  8487.3        & 8487.53    \\
$04^2\,E$    &  9131            &  9096.6        & 9112.90    \\
$13^1\,E$    &  9736            &  9649.2        & 9653.42    \\
$04^4\,E$    &  9802            &  9999.2        & 9996.72    \\
$22^2\,E$    &  10677        &  10646.        & 10644.59    \\
$05^1\,E$    &  10916        &  10827.        & 10862.46    \\
$31^1\,E$    &  11265        &  11349.        & 11322.31    \\
$14^2\,E$    &  11739        &  11656.        & 11657.69    \\
$05^5\,E$    &              &  12078.        & 12078.43    \\
\hline
$03^3\,A_2$    &  7482            &  7493.2        &    7491.89\\
$13^3\,A_2$    &  10243        &  10209.7        &    10209.55\\
\hline
\end{tabular}
\vspace{0.3cm}
\caption{Accuracy test of the adiabatic hyperspherical approximation and the improved coupled-channels hyperspherical calculation adopted for the computations presented in this paper.  Specifically, this table compares several vibrational energies in cm$^{-1}$ calculated in the present approach with the older adiabatic approximation results and those taken from a full three-dimensional diagonalization \cite{jaquet98}.}
\label{table:vibr_energies}
\end{table}

\section{Calculation of the raw DR rate coefficient and its average}
\label{sec:averaging}

The rate coefficient for dissociative recombination depends on the initial rovibrational state  $|rv\rangle$ of the ion. Using the defect of unitarity of the physical scattering matrix ${\cal S}^{phys}$, the DR rate coefficient $\alpha_{rv}$ for a particular rovibrational state $|rv\rangle$ is given by \cite{kokoouline03b}
\begin{eqnarray}
\label{eq:CS}
\alpha_{rv} (E_{el})=\frac{\pi}{\sqrt{2E_{el}}}\sum_{N} \frac{2N+1}{2N^++1}\left(1-\sum_{\substack{i=1,N_o}}{\cal S}_{i, i'}^{phys}(E_{el}){\cal S}_{i', i}^{\dagger phys}(E_{el})\right)\,,
\end{eqnarray}
where $E_{el}$ is the kinetic energy of the electron at infinity; the channel index $i'$ at  ${\cal S}$ corresponds to the initial  $|rv\rangle$ state. The scattering matrix ${\cal S}$ is calculated separately for each total angular momentum $N$ of the ion-electron system, $N^+$ is the angular momentum of the initial $|rv\rangle$ state. Notice that several values of $N$ may contribute to the DR rate coefficient for the given state $|rv\rangle$.

We now address the way we account for the experimental conditions in storage ring experiments, especially the experimental distribution over relative velocities of the ion and electron. In the storage ring experiments, the distribution is not uniform: the parallel component $u_{\parallel}$ of the $e^--$H$_3^+$ relative velocity $\vec u$ has a smaller distribution width than the perpendicular component $u_{\perp}$. The rate coefficient $\alpha_{rv} (E_{el})$ also depends on the rotational level.  In the experiments, the initial vibrational state is usually the ground state $\{00^0\}$, but several rotational levels are typically populated.  The population of different rotational levels is accounted for by introducing a finite rotational temperature $T_{rv}$ of H$_3^+$, though it should be remembered that this assumption that the ions are in thermodynamic equilibrium at some $T$ has not been explicitly confirmed experimentally. Present generation storage ring experiments measure the DR rate coefficient as a function of the parallel component $E_{\parallel}$ of the total relative energy $E_{el}$ of the ion and electron. The average over the non-uniform electron velocity distribution is then given by the following formula \cite{kokoouline05}
\begin{eqnarray}  \label{eq:averaging_final}
\alpha_{sr}(E_\parallel)=\frac{1}{N_{sr}}\sum_{rv}\int_{-\infty}^{\infty} du_{\parallel}\int_{0}^{\infty} dE_\perp \alpha_{rv}\left[(v_{\parallel}+u_{\parallel})^2/2+E_\perp\right]w_{sr}(rv,T_{rv})\,,
\end{eqnarray}
where the normalization constant $N_{sr}$ and the statistical factor $w_{sr}(rv,T_{rv})$ are 
\begin{eqnarray}
\label{eq:Weights_sr}
N_{sr}=\sum_{rv}\int_{-\infty}^{\infty} du_{||}\int_{0}^{\infty} dE_\perp w_{sr}(rv,T_{rv})\,,\nonumber\\
w_{sr}(rv,T_{rv})=(2I+1)(2N^++1)\exp\left(-\frac{E_{rv}}{kT_{rv}}\right)\exp\left(-\frac{u_{||}^2}{2\Delta E_{\parallel}}\right)\exp\left(-\frac{E_\perp}{\Delta E_{\perp}}\right)\,.
\end{eqnarray}
In the above equations, $\Delta E_{\perp}$ and $\Delta E_{||}$ are distribution widths (measured in energy units) for the parallel $\vec u_{\parallel}$ and perpendicular $\vec u_{\perp}$ components of the relative velocity $\vec v=\vec v_{\parallel}+\vec u_{\parallel}+\vec u_{\perp}$; $\vec v_{\parallel}$ represents the center of the velocity distribution, i.e. velocity at which the actual measurements are made in the storage ring experiments: $E_{\parallel}=v_{\parallel}^2/2$.  The perpendicular component of the energy is $E_{\perp}=u_{\perp}^2/2$. $I$ is the total nuclear spin, which can be $\frac{1}{2}$ or $\frac{3}{2}$ depending on the rovibrational state $|rv\rangle$; $E_{rv}$ is the energy of the H$_3^+$ rotational state (assuming that the vibrational state is always the same). The sums in the above equations are over all possible rovibrational states $|rv\rangle$ including all symmetries of the rovibrational states that can be populated at a given rotational temperature $T_{rv}$.

To compare with the storage ring experiments, one also must take into account the so-called toroidal effect, which is due to the geometry of the merged electron and ion beams, as there are two regions where the electrons are bent into or out of a trajectory that is parallel to the ions. The experimentally-observed rate coefficient with the toroidal effect correction is \cite{kokoouline05}:
\begin{eqnarray}
\label{eq:Toroidal_Correction}
\alpha_{tor}(E_{\parallel})=\alpha_{sr}(E_{\parallel}) + \frac{2}{L}\int_{0}^{l_{bend}}\alpha_{sr}(\tilde{E}_{\parallel}(x))dx\,.
\end{eqnarray}
The function $\tilde{E}_{\parallel}(x)$ and the length $l_{bend}$ account for the geometry of the merged electron and ion beams. A detailed discussion can be found in Ref. \cite{kokoouline05}. After the averaging procedures described above, the DR rate coefficient $\alpha_{tor}(E_{\parallel})$ can be compared with the raw experimental data from the storage ring experiments \cite{mccall03,mccall04,kreckel05}.

The thermally averaged DR rate coefficient relevant to a situation in which the ions and electrons are in common thermal equilibrium at temperature $T$ is calculated from the following integral:
\begin{eqnarray}
\label{eq:Thermal Rate}
\alpha_{th}(kT)=\frac{1}{N_{th}}\int_{0}^{\infty}\sum_{rv}\alpha_{rv}(E_{el})w(rv,kT)\sqrt{E_{el}}dE_{el}
\end{eqnarray}
where the normalization constant $N_{th}$ and the statistical factor $w_{th}(rv,kT)$ are 
\begin{eqnarray}
\label{eq:Weights}
N_{th}=\int_{0}^{\infty}\sum_{rv}w_{th}(rv,kT)\sqrt{E_{el}}dE_{el},\nonumber\\
w_{th}(rv,T)=(2I+1)(2N^+ +1)e^{-E_{rv}(R)/kT}e^{-E_{el}/kT}\,.
\end{eqnarray}

\section{Results}
\label{sec:results}

\begin{figure}[h]
\includegraphics[width=12cm]{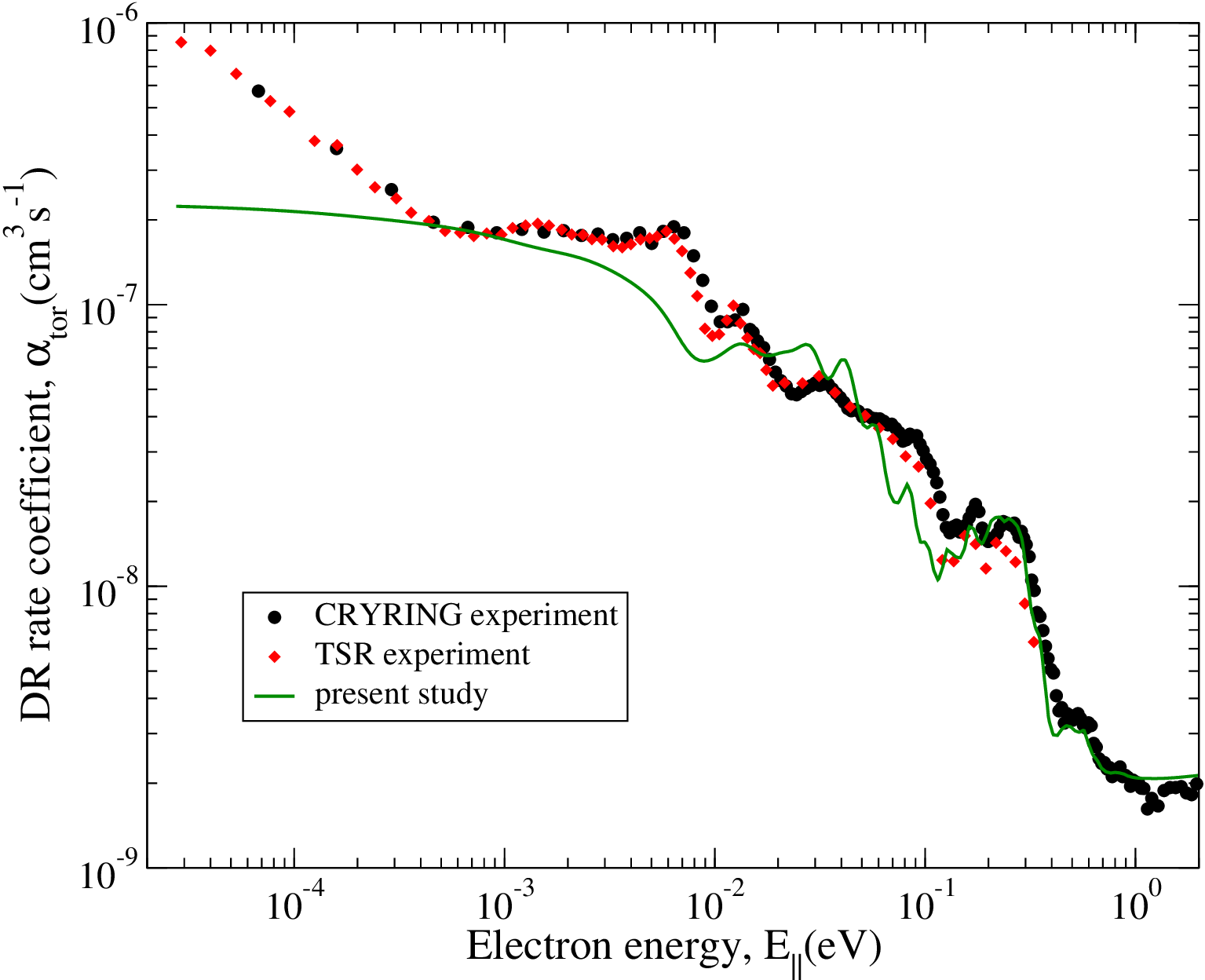}
\caption{\label{fig:alpha_SR} Comparison of experimental  \cite{mccall03,mccall04,kreckel05} (black circles and red diamonds) and present theoretical (solid line) dissociative recombination rate coefficients. In the theoretical calculation, the rotational temperature is $T_{rv}=$1000K, the widths $\Delta E_{\parallel}$ of the parallel component of the electron energy is 0.1 meV, and $\Delta E_{\perp}=$2 meV.}
\end{figure}

\begin{figure}[h]
\includegraphics[width=12cm]{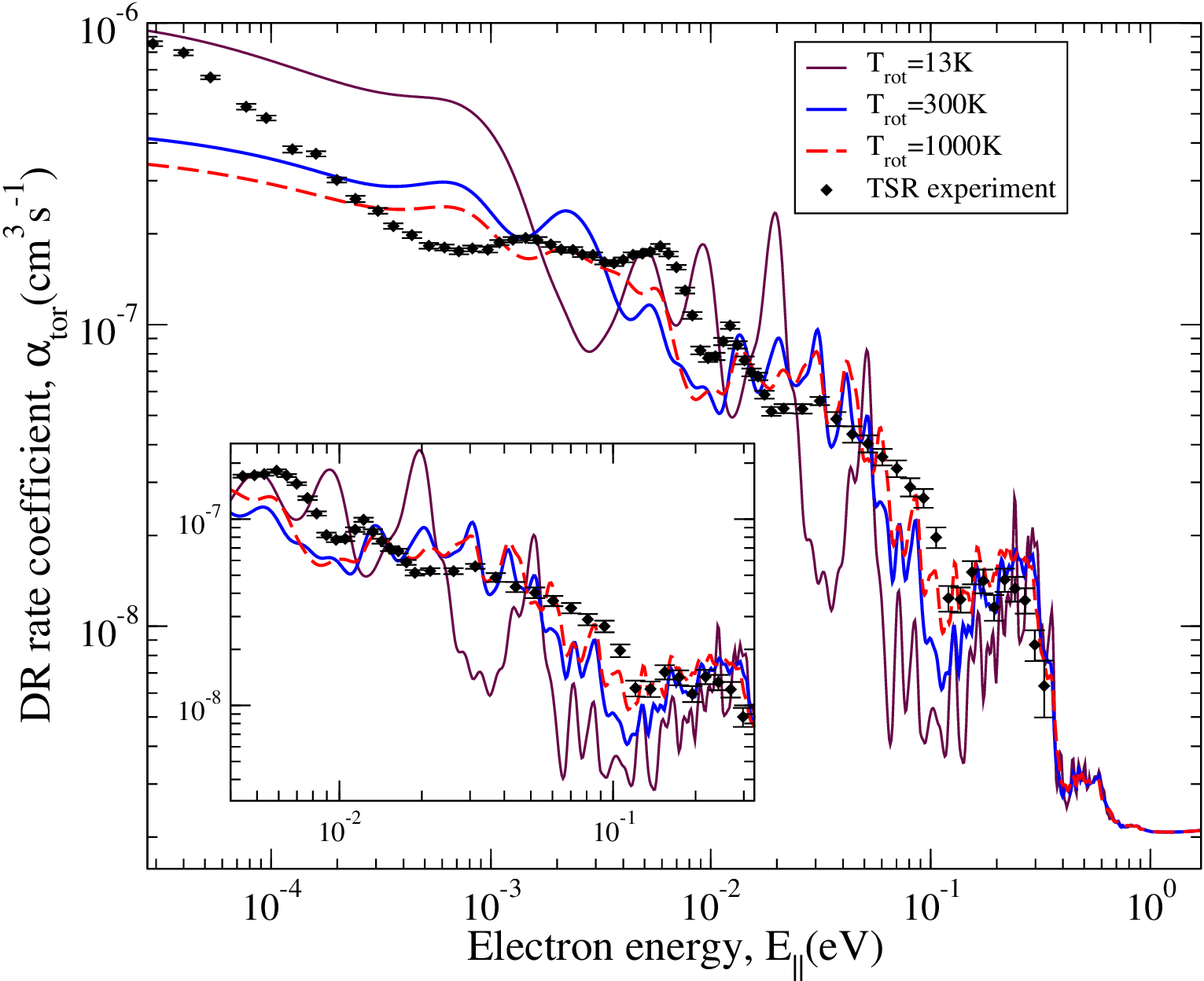}
\caption{\label{fig:alpha_SR_dif_rot} This figure presents calculations with different rotational temperatures in Eq.\eqref{eq:Weights_sr}. The theoretical DR rate coefficients obtained at higher rotational temperatures, e.g. 300K-1000K, agree better with both the CRYRING and TSR experiments.}
\end{figure}

\begin{figure}[h]
\includegraphics[width=12cm]{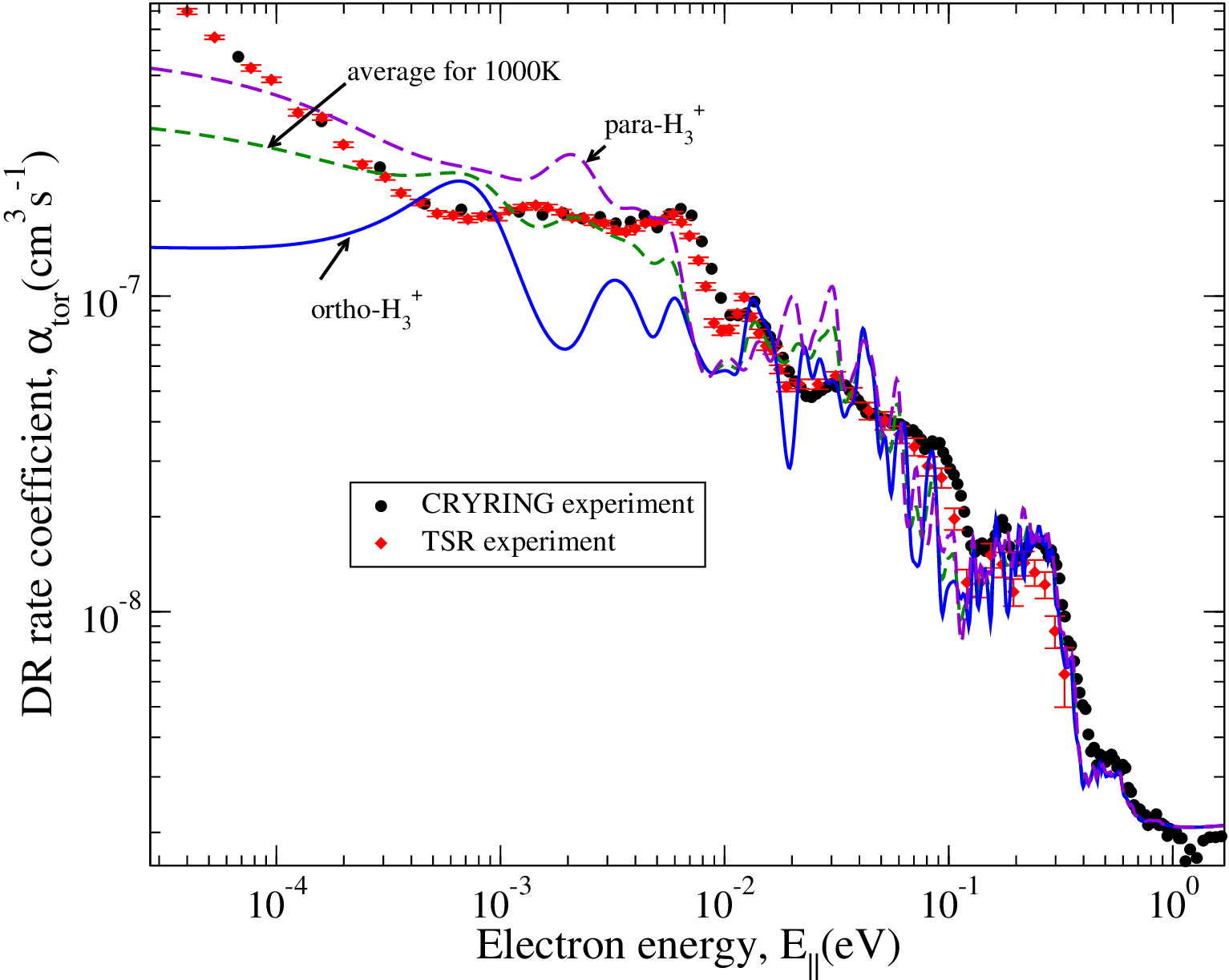}
\caption{\label{fig:alpha_ortho_para_new_exp} This figure compares the theoretical DR rate coefficient to the high-resolution storage ring experiment of Kreckel {\it et al.} \cite{kreckel05} carried out at TSR. The experimental resolution parameters are $\Delta E_{\parallel}$ and $\Delta E_{\perp}$ are 25$\mu$eV and 0.5 meV respectively. The theoretical curve shown has been calculated with these parameters and rotational temperature $T_{rv}=$1000~K. The figure also shows the theoretical DR rate coefficients calculated separately for ortho- and para- configurations of H$_3^+$ with the same parameters $\Delta E_{\parallel}$, $\Delta E_{\perp}$, and $T_{rv}$.}
\end{figure}

\begin{figure}[h]
\includegraphics[width=12cm]{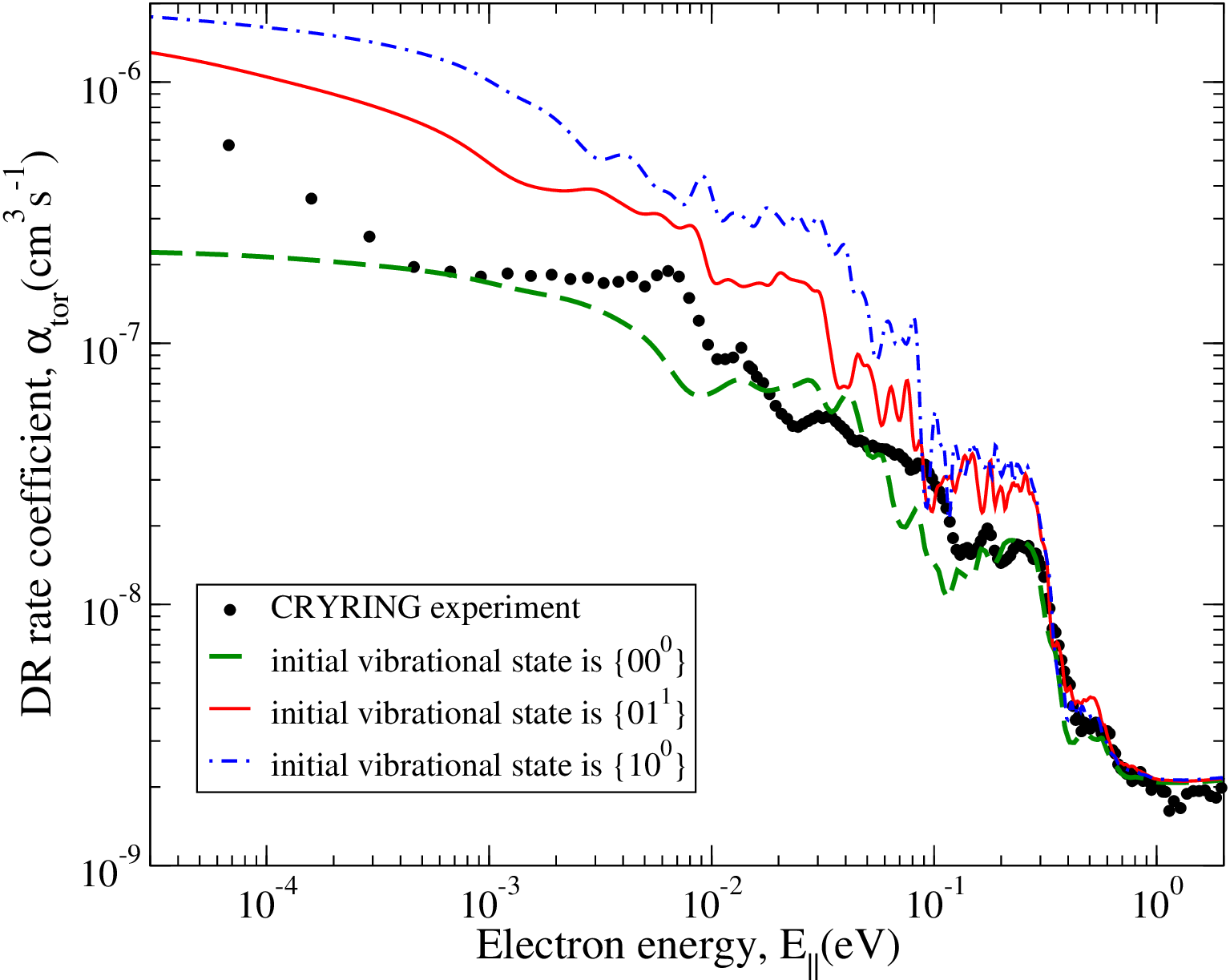}
\caption{\label{fig:alpha_SR_dif_vib} Comparison between the theoretical DR rate coefficients for the H$_3^+$ ion prepared in the ground and excited vibrational levels. At low energy, the DR rate coefficient for the vibrationally excited ion is significantly larger than for the ion in the ground state. This result is consistent with trends observed for DR rate coefficients in diatomic ions, where the DR rate coefficient typically increases with vibrational excitation. In the legend, the numbers in parentheses have the same meaning as in Fig. \ref{fig:alpha_SR_dif_rot}.}
\end{figure}

Figures \ref{fig:alpha_SR}, \ref{fig:alpha_SR_dif_rot}, \ref{fig:alpha_ortho_para_new_exp}, and \ref{fig:alpha_SR_dif_vib} summarize the computational results of the present study. Fig. \ref{fig:alpha_SR} compares the experimental DR rate coefficient from a recent storage ring experiment \cite{mccall03,mccall04} with the present calculation. Overall agreement with experiment is better than in the previous theory (see Fig. 6 in Ref. \cite{kokoouline05}). This is due to two factors: In the present treatment, we use more accurate vibrational wave functions, which are calculated using SVD, i.e. without the hyperspherical adiabatic approximation. The second improvement is due to the larger rotational temperature $T_{rv}$ and the larger number of rotational states that are taken into account in the averaging formula Eq. (\ref{eq:averaging_final}).
The energy of the highest rotational level (5,1) included in the present calculation is 1250.3 cm$^{-1}$ \cite{lindsay01}. The energy of the lowest state (1,1) allowed for H$_3^+$ is 64.1 cm$^{-1}$ above the symmetry-forbidden state (0,0). In Table \ref{table:rot_energies} we show the partial contributions of the rotational states with different $N^+$ to the total DR rate coefficient with $T_{rv}$=300~K for four different energies 0.001, 0.01, 0.1 and 1 eV. Although the relative population of the $N^+=5$ rotational states is about 3\% at $T_{rv}$=300~K, there can be accidental cases, at certain energies where other contributions happen to be small, where it can become an important contributor to the observed DR rate. For example, at a collision energy 0.0997 eV, the cumulative contribution of the states with $N^+=5$ contributes 14\% of the calculated DR rate.

\begin{table}
\vspace{0.3cm}
\begin{tabular}{|p{3.cm}|p{4.1cm}|p{1.5cm}|p{1.5cm}|p{1.5cm}|p{1.5cm}|p{1.5cm}|}
\hline 
Energy (eV)    &Total DR rate coefficient (cm$^{3}$/s)    &$N^{+}=1$   &$N^{+}=2$  &$N^{+}=3$  &$N^{+}=4$ &$N^{+}=5$   \\
\hline
 0.00101            & $2.19\times 10^{-7}$                     &  0.295      & 0.198     & 0.364     & 0.126    & 0.017    \\
 0.0103             & $4.79\times 10^{-8}$                     &  0.388      & 0.139     & 0.301     & 0.136    & 0.035    \\
 0.0997             & $6.30\times 10^{-9}$                     &  0.157      & 0.238     & 0.266     & 0.196    & 0.143    \\
 1.02               & $7.10\times 10^{-11}$                    &  0.324      & 0.201     & 0.315     & 0.111    & 0.049    \\
\hline
\end{tabular}
\vspace{0.3cm}
\caption{Partial fractional contributions to the DR rate coefficient from individual ionic angular momenta, $N^{+}=1,\cdots ,5$, at four different energies. The DR rate coefficient is calculated for $T_{rv}$=300K. In the table, the DR rate coefficient and partial contributions are calculated without the toroidal averaging. For larger temperatures, higher rotational states can  have important contributions at certain electron energies.}
\label{table:rot_energies}
\end{table}

Figure \ref{fig:alpha_SR_dif_rot} demonstrates the theoretical DR rate coefficients we obtain for different ionic rotational temperatures $T_{rv}$. The results with higher $T_{rv}$ agree better with the experiments than when the experimentally estimated rotational temperature is adopted. While this better agreement at a higher rotational temperature could be fortuitous, our results suggest that it is worth exploring whether the rotational temperature in both of the recent storage-ring experiments \cite{mccall03,mccall04,kreckel05} might be larger than 40~K or 13~K, respectively. (40~K and 13~K are the estimated rotational temperatures in the two experiments). This conclusion would conflict with another suggestion by Kreckel {\it et al.} \cite{kreckel05}, that only the two lowest rotational states are populated in the recent storage ring experiments \cite{mccall03,mccall04,kreckel05} and that, therefore, the rotational temperature $T_{rv}$ is about 13~K---40~K. In fact, it was demonstrated previously, that if the electron energy is high enough, the electron-ion collision might not only cause DR or result in an elastic collision, but it can also result in rotational excitation of the ions when they circulate in the storage ring \cite{kokoouline03b,faure06}. In all the tests we have carried out for temperatures in the range 13~K-40~K, the calculated DR rate coefficient has pronounced structure due to Rydberg states present in the raw rate coefficient of Eq. \ref{eq:CS}. The experimental DR rate coefficient has some structure, but it is less pronounced than our averaged and convolved theoretical rate coefficient. Since the resonances due to the Rydberg states are smeared out in the experiments, it suggests the possibility that in the experiment there could be an additional source of broadening. The broadening could arise from a higher rotational temperature, from a broadened electron energy distribution, from additional broadening associated with the toroidal region, or perhaps from something else. 

An alternative possibility that cannot be ruled out is that our theoretical treatment might have underestimated the resonance widths.  For the resonances that dominate the DR rate, the predissociation partial width is larger than the autoionization partial width, and under those conditions, the calculated DR rate is comparatively insensitive to changes in the predissociation linewidth.  Thus, it would be a valuable benchmark for experiments (or other, improved theories) to determine the predissociation partial widths of individual resonances above the ionization threshold, to provide a direct test of the accuracy of our present calculations at the level of spectroscopic accuracy.

One possible source of rotational excitation could be the repeated circulation of the molecular ions through electron cooler during the ramping of the cathode voltage \cite{mccall04}. The authors of Ref. \cite{mccall04} deduced that the rotational temperature is 40 K based on theoretical cross-sections of the rotational excitation of H$_3^+$ given in Ref. \cite{faure02}. Since then, the cross-sections have been reconsidered and corrected \cite{faure06}. Correspondingly, we revisit here the arguments of Ref. \cite{mccall04} based on the new inelastic probabilities determined by Ref. \cite{faure06}.  If we take the (1,1)$\to$(2,1) rotational excitation cross-section to be 710 \AA$^2$ from Ref. \cite{faure06} instead of 210 \AA$^2$ from Ref. \cite{faure02}, we obtain the relative population 7.5\% instead of  2.2\% in Ref. \cite{faure02}. If we take the (1,0)$\to$(3,0) rotational excitation cross-section to be 270 \AA$^2$ from Ref. \cite{faure06} instead of 120 \AA$^2$ from Ref. \cite{faure02}, we obtain the relative population 4.6\% instead of 2.0\% in Ref. \cite{faure02}. The relative population 7.5\% of the (2,1) states corresponds to the temperature $T_{rv}$=96~K, the relative population  4.6\% of the (3,0) states gives $T_{rv}$=210~K. Both values of the temperature are significantly higher than the values (40~K and 13~K) quoted in the experimental papers, although not as high as 1000K, in our present estimation.

The disagreement between the current theory and experimental data around energy $E_{\parallel}=$~6~meV (see Fig. \ref{fig:alpha_SR}) could be caused by errors in the calculated positions of Rydberg states in that region, which are attached to highly-excited rovibrational levels of the ion. In our calculation, we use energy-independent quantum defects even though they depend weakly on the principal quantum number. In addition, the accuracy of calculation of energies for the highly-excited rovibrational levels may be of the order of 6~meV. Note that there is also a disagreement with experiment in the region of very small energies, below 0.2 meV. This region is well below the net effective energy resolution of the experiment ($\Delta E_{\perp}=$2 meV). The experimental rate coefficient appears to behave there as ${E_{\parallel}}^{-1/2}$, which is the same total energy dependence expected for the raw, unconvolved DR rate coefficient. However, in our theoretical calculation, we of course included the convolution according to the perpendicular energy distribution with the width $\Delta E_{\perp}=$2 meV. This makes the convolved theoretical DR rate coefficient become essentially flat at very low energy $E_{\parallel} \ll 2$ meV even though the raw theoretical rate coefficient also grows as $E^{-1/2}$ at very low energy. This suggests that in the experiment, the distribution of relative electron energies could be even more complicated than discussed above, and in particular, the resolution might be even better at very low energies than the quoted energy resolution.
Another possible explanation for the disagreement is that a Rydberg resonance exists in H$_3$ at an energy just above (+0.3 meV) the (1,1) rotational state of the ion. The predissociation linewidth of the resonance must also be of the order 0.3 meV. We have made a simple test calculation in which we artificially tuned one of the Rydberg resonance to be placed just above the (1,1) ionization threshold. The resulting theoretical DR rate coefficient looks very similar to the experimental DR rate coefficients at energies below 0.5 meV. Therefore, the sharp increase of the experimental DR rate coefficient for energies below 0.3 meV requires additional consideration.  Similar discrepancies have been observed between DR theory and experiment at very low parallel energies, in other systems such as LiH$^+$,\cite{CurikGreene2007} so this could be a systematic issue for theory and experiment to confront which extends beyond the H$_3^+$ system alone.

Figure \ref{fig:alpha_ortho_para_new_exp} compares the present theoretical DR rate coefficient (dashed grey curve) with the recent TSR storage ring experiment by Kreckel {\it et al.} \cite{kreckel05}. In the TSR experiment, the parallel and perpendicular energy resolution parameters $\Delta E_{\parallel}=$25$\mu$eV and $\Delta E_{\perp}$=0.5 meV  are slightly smaller than in the CRYRING experiment \cite{mccall03,mccall04}. Therefore, the theoretical curve shown in the figure has been correspondingly convolved using these parameters. These calculations have also assumed that the target ion rotational temperature $T_{rv}=$ is equal to 1000~K. The overall agreement between theory and experiment is good except in the energy region below 0.15 meV already discussed above. As one can see, when the width $\Delta E_{\perp}$ is decreased from 2 meV in Fig. \ref{fig:alpha_SR} to 0.5 meV in Fig. \ref{fig:alpha_ortho_para_new_exp}, the theoretical rate coefficient agrees better with the sharp increase of the experimental DR rate coefficient for energies below 0.5 meV. 

In the TSR experiment, Kreckel {\it et al.} \cite{kreckel05}. observed the dependence of the DR rate coefficient on the nuclear spin of H$_3^+$. They found that para-H$_3^+$ has a larger DR rate coefficient than ortho-H$_3^+$ for low energies ($<0.5$ meV). The previous theory \cite{kokoouline03b} has predicted different DR rate coefficient for para-H$_3^+$ and ortho-H$_3^+$: At low energies the theoretical DR rate coefficient for ortho-H$_3^+$ was larger than for para-H$_3^+$, i.e. opposite to what was observed in the experiment by Kreckel {\it et al.}  We now revisit this issue in the context of our new and presumably improved theoretical description. Figure \ref{fig:alpha_ortho_para_new_exp} shows the separate ortho-H$_3^+$ and para-H$_3^+$ DR rate coefficients calculated in the present treatment. The para-H$_3^+$ rate coefficient is significantly higher than the rate coefficient obtained for ortho-H$_3^+$. This dramatic difference between the ortho-H$_3^+$ and para-H$_3^+$ rate coefficients obtained at low electron energies and those of our previous theoretical study appears to result from slightly different positions of the calculated Rydberg H$_3$ states whose energies lie close to the (1,1) and (1,0) ionic rotational states.

Figure \ref{fig:alpha_SR_dif_vib} presents our theoretical DR rate coefficients obtained for a target ion that is initially in an excited vibrational state. These calculations have been carried out for the first $\{01^1\}$ and second $\{10^0\}$ excited ionic vibrational states. (In \{\} we specify the vibrational quantum numbers of the ion using the normal mode notation \cite{lindsay01}.) The energy of the lowest rotational state (0,0) for $\{01^1\}$ is 2521.4 cm$^{-1}$, and the energy of the lowest rotational state (1,1) for $\{10^0\}$ is 3240.7 cm$^{-1}$ \cite{lindsay01}. The DR rate coefficient for the two excited vibrational levels of the ion is higher than the DR rate coefficient for the ground state, which is reasonable, considering that a similar qualitative increase of the DR rate coefficient was previously observed in both theory and experiment for diatomic ions.

\begin{figure}[h]
\includegraphics[width=12cm]{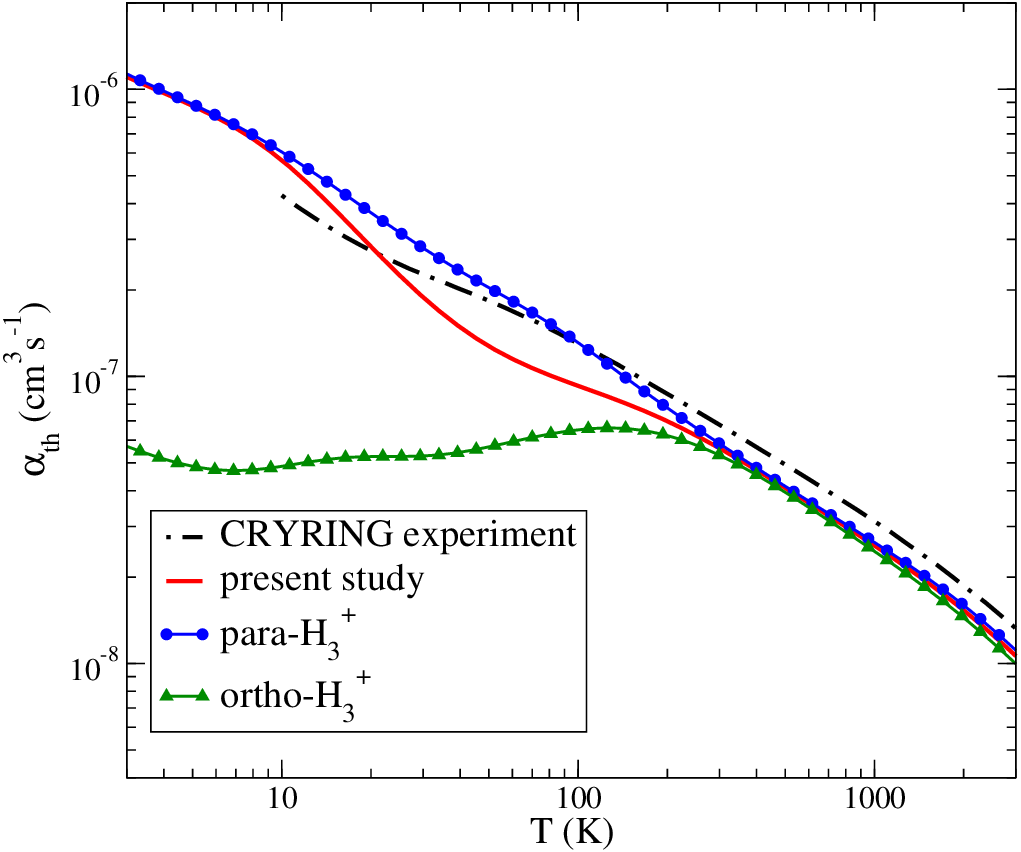}
\caption{\label{fig:ther_rate} The present theoretical thermal rate coefficient for dissociative recombination of H$_3^+$ is compared with the experimental rate coefficient deduced from the storage ring experiment of McCall {\it et al.}\cite{mccall03,mccall04}} 
\end{figure}

Finally, Fig. \ref{fig:ther_rate} compares the theoretical and experimental thermal rate coefficients. The theoretical rate coefficients are obtained directly from the raw theoretical data using Eqs. (\ref{eq:Thermal Rate}) and (\ref{eq:Weights}). Thus, the toroidal effect and the finite widths  $\Delta E_{\perp}$, $\Delta E_{\parallel}$, $kT_{rv}$ are not present in these theoretical results. In fact, our calculation shows that the inclusion of the toroidal correction increases the thermal rate coefficient by about 20 \% approximately uniformly for all energies. Thus, the agreement with experiment is good. The theoretical thermal rate coefficient at $300$K is $5.6\times 10^{-8}$ cm$^3$/s. 

\section{Conclusions}
\label{sec:conclusion}

In summary, we would like to emphasize the following results from the present study:

We have calculated the rate coefficient of H$_3^+$ dissociative recombination using an improved description of the ionic vibrational states, and including more target rotational states. The resulting theoretical rate coefficient agrees reasonably well with two recent storage ring experiments \cite{mccall03,mccall04,kreckel05}. The agreement with experiment has been improved over that achieved in the previous theoretical study \cite{kokoouline03b,kokoouline03a}; in particular, the present results may point to a resolution of the largest previous discrepancy, in the energy range from 0.04 eV to 0.15 eV. 
However, the improved agreement with the experimental data in that energy range was only obtained when we assume that the rotational temperature $T_{rv}$ of H$_3^+$ is significantly larger than 40~K and 13~K, the values given in the experimental study \cite{mccall03,mccall04,kreckel05}. Since no direct measurement of the temperature has actually been made inside the storage ring in those experiments (except indirectly for zero-energy collisions), there is a possibility that the ions get rotationally excited before the DR rate coefficient measurements are conducted. 
It was shown previously that the probability of rotational excitation of the ion by electrons is comparable to or larger than the DR probability, at energies where rotational excitation is energetically allowed. Thus, the rotational temperature in the experiment could be larger than 40 K or the ions might not even be in thermal equilibrium at any temperature.  Thus, it would be desirable to monitor the rotational temperature during the DR measurement. Another possibile way to explore this effect is to artificially increase the temperature of the electron cooler or the width $\Delta E_\perp$ during the DR measurements and ascertain the temperature at which the DR rate coefficient starts to become sensitive to the temperature.

We have calculated the DR rate coefficients for separate ortho- and para-configurations of H$_3^+$. At energies below 10 meV, the DR rate coefficient for para-H$_3^+$ is an order of magnitude larger than for ortho-H$_3^+$. The experiment also shows that the para-H$_3^+$  DR rate coefficient is larger. However, since the ortho-/para-ratio in the experiment is not known it is not clear what is experimental DR rate coefficients for pure para-H$_3^+$ and ortho-H$_3^+$.  Our previous calculations stressed that the ortho-para ratio of DR rates at very low collision energies should be used with some caution, because the rates at energies below 100K begin to get very sensitive to the specific resonance positions at the meV level. Those cautionary remarks are still applicable to the present results for the ratio of ortho and para DR rates at low energy.  However, if the present order-of-magnitude difference of the low energy DR rate survives future improvements in theory and experiment, it will be interesting to explore possible implications of this difference for the chemistry of interstellar clouds.

Finally, the calculated DR rate coefficient for ions prepared in excited vibrational states is larger than in the ground vibrational state. Our new theoretical values for the DR rate coefficients of excited vibrational states will hopefully be tested one day in a storage-ring experiment. Currently, experimental DR measurements are made after the ions have been cooled. In principle, it seems possible to carry out this measurement using vibrationally-hot ions and, to monitor the DR rate coefficient as a function of the vibrational temperature. Such an experiment might give deeper insights into the energetics and target state dependence of the DR process, and it could then be compared with the present theoretical predictions.

\begin{acknowledgments}
We particularly thank A. Wolf, H. Kreckel, A. Petrignani for extensive and helpful discussions. We have also benefitted from the ongoing use of B. Esry's 3-body code that solves the adiabatic hyperspherical eigenvalue problem.  This work has been supported by the National Science Foundation under Grant No. PHY-0427460 and Grant No. PHY-0427376, by an allocation of NERSC and NCSA (project \# PHY-040022) supercomputing resources.The work of CHG has also been partially supported by the Miller Institute for Basic Research in Science, and from the Alexander von Humboldt Foundation.
\end{acknowledgments}

\end{document}